# Pulsed Laser Template Engineering- PLATEN

Dhiman Biswas[1,2], Junyeob Song[3], Francisco Guzman[4], Levi Brown[4], Yiwei Ju[4], Nisha Geng[5], Pralay Paul[1,2], Sumit Goswami[6], Casey Kerr[1,2], Sreehari Puthan Purayil[1,2], Ben Summers[1,7], Preston Larson[8], Binbin Weng[6,7], Bin Wang[5,10], Horst Hahn[9], Xiaoxing Pan[4], Alisa Javadi[1,7], Henri Lezec[3], and Thirumalai Venkatesan[1,2,3,7]

[1]Center for Quantum Technology and Research, University of Oklahoma OK 73019

[2]Department of Physics and Astronomy, University of Oklahoma OK 73019

[3]NIST, Gaithersburg, MD

[4]Department of Materials Science, UC Irvine, Riverside, CA

[5]School of Sustainable Chemical, Biological and Materials Engineering, University of Oklahoma OK 73019

[6]MREC, University of Oklahoma OK 73019

[7]School of ECE, University of Oklahoma OK 73019

[8]SRNML, University of Oklahoma OK 73019

[9]Department of Materials Science and Engineering, University of Arizona, Tucson AZ 85721

[10]Department of Chemical and Biological Engineering, Tufts University, Medford, MA 02155

## Abstract

Thin films of functional inorganic materials, particularly oxides, play a vital role in opto-electronics, enabling applications that range from active optical components to MEMS-based architectures. Achieving high aspect ratio patterning of these functional materials remains a significant challenge, as many of their constituent elements do not readily form volatile compounds required for conventional reactive ion etch processes. We introduce a novel approach, **P**ulsed **La**ser **T**emplate **EN**gineering (PLATEN), which offers a more accessible route for patterning materials that are typically difficult to etch. This technique involves depositing functional films using the **P**ulsed **L**aser **D**eposition (PLD) process onto silicon substrates that have been pre-patterned using reactive ion etching to create high aspect ratio features. Due to the highly forward-directed nature of the PLD process, the deposited films replicate closely the topography of the patterned silicon, without coatings the sidewalls. This process remains effective even at feature sizes down to approximately 50 nm. The oxide films replicate the underlying silicon pattern to a thickness of 80 nm. For thickness beyond 80 nm the patterns develop a waist at the midpoint which scales with film thickness and is not dependent on the feature size. In this paper, we present a detailed analysis of the PLATEN process, including deviations from ideal pattern replication in sub-micron features as a function of film thickness, and demonstrate near single crystalline growth of oxides on the patterned silicon substrate, demonstrating the potential of PLATEN technique for active opto-electronic materials.

**Introduction:** Heterogeneous integration [1] [2] [3] [4] [5] [6] is the next evolving chapter in semiconductor technology (beyond silicon) which is currently moving on to the creation of microsystems on a silicon platform. Examples of such microsystems range from optical interconnect sub-systems to small drones capable of payloads with a small camera, micro-satellites with a variety of on-board instruments and so on [7] [8]. Functional oxide materials (ferro-electric, piezo-electric, electro-optic, magneto-optic, ferromagnetic, etc.) need to be integrated on silicon surfaces to provide the required capabilities. In many cases patterning of such films on silicon is necessary [9] [10] [11]. For example, in the context of active optical circuitry on silicon, materials like $LiNbO_3$, $BaTiO_3$ and $LiTaO_3$ are notoriously difficult to etch using conventional reactive ion etching techniques, as they do not readily form volatile compounds [12] [13] [14]. Dry etching of functional oxides like $LiNbO_3$, $LiTaO_3$, $BaTiO_3$, by fluorine-based plasma ($CHF_3$, $CF_4$, $SF_6$) form nonvolatile species like $LiF$, $TiF_4$ and $BaF_2$ which redeposit on the sample and acts as micro masks, causing sidewall roughness and significant increase in etching time. Furthermore, those nonvolatile metal halides along with polymers used affect the chamber and cause flaking and particle contamination and hence excessive downtime for cleaning. The pioneering work of Pearton et al. addresses these challenges [15]. As a result, these materials are generally ion milled, which results in considerable sidewall roughness. Culmination of all these makes the process of etching functional oxides more complicated and resource intensive. To circumvent this problem, we propose here a simple technique, **P**ulsed **La**ser **T**emplate **EN**gineering (PLATEN), which enables the creation of structures down to tens of nm scale using the forward directed nature of the **P**ulsed **L**aser **D**eposition (PLD) process onto pre-patterned silicon substrates. In the PLD process, the 248 nm laser pulses ablate the material from the target surface and create a plasma plume containing the ions and neutral fragments of the target material. Beside preserving the stoichiometry, this plasma plume is highly forward directed with an angular dependence of $sin^m\theta$, where m is greater than 8 [16]. This forward directed plasma is expected to align the deposited material along the pattern on the silicon with minimal sidewall coating. This paper discusses the details of this technique and the evolution of the self-aligned patterned deposit as a function of thickness, deposition temperature and crystallinity.

**Experiment:**

The PLATEN process comprises several sequential steps. Initially, silicon is patterned to form high aspect ratio structures. Subsequently, a thin buffer layer of a crystalline film, such as YSZ, is deposited onto the patterned structures to facilitate the epitaxial growth of functional oxides such as $LiNbO_3$, $CeO_2$ or $BaTiO_3$ (Fig. 1(a)).

The growth of the buffer layer is followed by the deposition of the functional oxide, which replicates the underlying patterned silicon structures. Each of these experimental steps is discussed in detail below. To clarify the rationale behind the selection of different materials, Table 1 summarizes the materials investigated, their growth characteristics, and their relevance to the overall process.

**Table 1**

| Material | Short form | Functionality | Growth Nature (>750C) |
|---|---|---|---|
| Y stabilized zirconia | YSZ | Epitaxial buffer layer on Si | Layer-by-layer growth |
| W doped Indium oxide | IWO | Transparent conducting oxide | Hybrid of Homogeneous nucleation and layer by layer growth. |
| Cerium oxide | $CeO_2$ | Quantum host for color centers | Layer-by-layer growth |

**High aspect ratio patterning of the silicon wafer:** Overall, we employed a lift-off process combined with electron beam lithography and inductively coupled plasma reactive ion etching (ICP-RIE) to create high aspect ratio patterns. A 4-inch silicon wafer was first cleaned using the RCA method, followed by spin-coating of a 350 nm-thick electron beam resist (ZEON ZEP520A) and baked at 180 °C for 10 minutes. After development in hexyl acetate, a 50 nm-thick $Al_2O_3$ layer was deposited by electron beam evaporation (Denton Vacuum Infinity 22). The resist was subsequently removed using a lift-off solvent (Kayaku Remover PG), leaving a patterned $Al_2O_3$ hard mask. ICP-RIE (Oxford Instruments PlasmaLab System 100) was then performed using a gas mixture of $SF_6$ and $C_4F_8$ (Bosch Process). The high anisotropy achieved during ICP-RIE arises from the independent control of plasma density and ion energy, which enables directional etching with minimal lateral erosion. In this process, $SF_6$ provides fluorine radicals for chemical etching of silicon, while $C_4F_8$ forms a fluorocarbon polymer that passivates the sidewalls. Sidewall roughness can be further minimized by optimizing the $SF_6$/ $C_4F_8$ flow ratio. After etching, the $Al_2O_3$ hard mask was removed using a mixture of $NH_4OH$ and $H_2O_2$.

**Deposition of buffer epitaxial oxide film:** The presence of a native silica layer on the silicon surface poses a significant obstacle to the epitaxial growth of functional oxides on silicon. There are many techniques to circumvent this issue, high temperature UHV anneal [17], Ar sputtering [18], fluorine base RIE [19] and, most commonly the use of buffered Hf (6:1 mixture of $NH_4F$ to HF) to strip off the oxide layer [20] prior to deposition of a buffer layer. Fortunately, the growth of epitaxial YSZ films on silicon (with a native oxide layer) using PLD is well studied, which doesn't require stripping off the thermal oxide layer [21]. The process begins by heating the substrate to 950 °C, a temperature sufficiently high to activate the mobility of zirconia ions and promote effective film growth. This first step is followed by the deposition of an ultra-thin (a few nm) seed layer under high vacuum of $9.9 \times 10^{-4}$ Pa. At such low oxygen pressures, the zirconia plume is oxygen deficient and because of the higher oxygen affinity of Zr it etches the silica off the silicon's surface by scavenging the oxygen off the silicon dioxide resulting in volatile SiO which depletes the surface oxide. The YSZ layer upon reaching the silicon surface begins to grow epitaxially due to the favorable lattice match with silicon forming a seed layer for further growth.

After the seed layer deposition, tens of nm of YSZ is deposited at 2 Hz under increased oxygen pressure (6.53X10$^{-2}$ Pa) to further grow the epitaxial buffer layer and to also reduce the oxygen vacancy defects in the seed layer. This epitaxial growth of YSZ on silicon enables the deposition of a wide range of functional oxides like $CeO_2$ (quantum hosts), MgO (buffer layer for (111) $BaTiO_3$ growth), ITO (transparent conducting oxide), $SrTiO_3$ (buffer for (100) BaTiO3 or a conducting electrode with Nb doping), $BaTiO_3$ (electro-optic layer) and so on.

To understand the PLATEN process different growth conditions have been studied:

1. Room temperature growth of amorphous oxide layers - IWO and YSZ.
2. High temperature growth of two different material systems:
   (a) material growth dominated by layer-by-layer growth- YSZ and $CeO_2$ and
   (b) material growth dominated by homogeneous nucleation - IWO.

**Room temperature PLATEN with IWO:** In this experiment, we deposited IWO layers on pre-patterned silicon with grating structures with grating widths of 500 nm and 1 micron pitch. The deposited pattern is shown in Fig. 1 (b). The deposited material on the top of the grating follows exactly the shape of the silicon. The material in between deposits at the bottom of the etched pits. There is no material deposited on the sidewalls, the oxides at the top and bottom of the trench are disconnected if the etch depth is larger than the deposited film thickness. If one looks at the edges of the bottom layers a V groove is seen which can only be explained by a shadowing effect of a widening pattern at the top. In Figure 1(b) the groove starts only after about 80 nm of material has been deposited. This suggests the following: initially, the pattern faithfully reproduces the shape of the silicon, but beyond approx. 80 nm it tends to develop a waist. A more detailed study was carried out as a function of the deposited thickness.

**Spatial Resolution of the PLATEN process:** In Fig. 2 is shown the PLATEN of $CeO_2$ films of 100 nm thickness deposited on patterned silicon structures ranging from 500 to below 50 nm. The film was deposited at a temperature of 800 C at an oxygen pressure of 1.5X10$^{-1}$ Pa. The $CeO_2$ faithfully reproduces the shape of the patterned silicon with a waist of about 20 nm which becomes more obvious when one looks at the structures on the silicon pillars with dimensions of sub-50nm (Fig. 2 (d)). Even though the film is grown at such high temperature since $CeO_2$ growth is dominated by layer-by-layer growth the surface is very smooth, and the side wall roughness mimics the roughness of the patterned silicon which for a good Bosch process can be brought down to sub-nm dimensions. So, the PLATEN idea works down to 50 nm lateral dimensions of the Si-substrate and possibly beyond, though the 50 nm silicon pillar resulted from an artifact during the RIE process. Currently, the spatial resolution of the PLATEN process is limited by our ability to create a finer silicon pattern.

**High temperature growth of two different material systems IWO and $CeO_2$:** To explore the effect of growth morphology (layer by layer vs homogenous nucleation), we compare the PLATEN process for $CeO_2$ and IWO which have layer-by-layer and homogeneous nucleation dominated growth processes, respectively. Unlike popular ITO (Tin doped Indium Oxide) which grows layer by layer on YSZ , the IWO tends to have significant homogenous nucleation component (Figure 2S(b)) and we attribute this to the larger ionic radius mismatch of tungsten (25%) vs lower mismatch of tin (13.7%) with Indium. Fig. 2S shows the cross sections for these two different materials. In the case of $CeO_2$ (Fig. 2S(a)) the film surface is smooth on the top surface, and the side walls have the same roughness as that of the patterned silicon below which is estimated to be of the order of nms. However, the surface of IWO (Fig. 2S(b)) is rough with crystallites showing spontaneous nucleation and the side walls of the IWO are also much rougher. So, for the PLATEN process to work for a crystalline layer the solid phase growth process must be dominated by layer-by-layer growth.

**Thickness dependence of the film growth:** We noticed that for both room and high temperature growth (for YSZ) the waist develops and increases with thickness qualitatively in a similar way. The waist is defined as the maximum deviation of the film edge from the patterned substrate's edge and is close to the mid-plane of the film thickness. However, for the high temperature growth the waist has sharper facets unlike the rounded waists for room temperature growth (Figs.1S, 4S and 4). We attribute this waist formation to the film's effort to minimize the surface energy. The waist forms close to the mid-plane of the film, corresponding to half of its total thickness (Figs. 3 and 4S). The top width of the film is always equal to the width of the grating.

The errors in the waist measurements are smaller for the high temperature deposition unlike that for the room temperature due to the formation of facets at the midpoint. However, the thickness dependence shows very similar features irrespective of the deposition temperature. The threshold thickness of about 80 nm for waist formation, the gradual increase in the waist width till a thickness of 350 nm and its gradual saturation for larger thicknesses are similar for both cases (Figs 3b). The relationship between waist width and the film thickness beyond 80 nm can be fitted with a simple second order polynomial function.

We make use of the fact that the waist does not form until a thickness of 80 nm is reached. So, it made us consider the possibility of using this to fabricate structures without any waist. A series of deposition (of YSZ at 950C and $6.53 \times 10^{-2}$ Pa) were made whereby a film of thickness around 214 nm was formed with five sequential depositions with a wait time of 15 minutes between the depositions. The cross section of the final film is shown in Fig. 3S in comparison with a similar thickness film done in a single step. In both cases an identical waist is clearly formed which exactly corresponds to the expected waist thickness and located at the midpoint of the film plane. This indicates that the waist formation is strongly influenced by the thickness of the film and possibly the shape of the pattern underneath (Figure 10S). Furthermore, the formation of the waist at room temperature for different thickness (Figure 3) also indicates the near temperature independence of waist formation process. A comparison of the waist formation on different feature sizes for 200, 100 and sub-50 nm cylinders shows that the waist is not strongly dependent on the feature size (Figure 10S). However, the waist for the cylindrical pattern is almost double that of the grating features implying possible shape dependence of the waist width. This requires further study.

**Crystallinity of the templated self-aligned layer:** Fig. 5 (a-d) shows a TEM cross section of a 200 nm $CeO_2$ deposited (at 800C, $1.5 \times 10^{-1}$ Pa) on a 24 nm thick YSZ buffer layer on a prepatterned silicon grating with a width of 500 nm and a micron pitch. This is compared with the same heterostructure deposited on an unpatterned silicon surface (Fig. 5S (e-h)). The heterostructure on unpatterned silicon exhibits single crystallinity throughout the layer while the ones on the patterned surface show polycrystalline domains. We attribute this to the interaction between the hard mask ($Al2O_x$) used for the RIE process and the potential damage caused by the mask to the pristine, atomically smooth silicon surface.

To circumvent this issue, we used a novel approach to this patterning technique. Before etching we deposit an ultra-thin (10-30 nm) layer of single crystal YSZ on silicon. The robustness of crystalline YSZ layer ensures that the surface of the sample doesn't get affected during the subsequent processing. After that, we use $BCl_3$ to etch the YSZ using RIE, followed by $SF_6$ reactive ion etch of the silicon underneath (Fig 1a) Using this technique, we created grating structures (500 µm) on silicon with crystalline YSZ coating. The grating area was chosen to maximize the XRD signal intensity from our sample at every stage of processing. PLD depositions of crystalline layers (YSZ, $CeO_2$) were done under optimum conditions followed by XRD and cross-sectional TEM study. The sample on which further YSZ was deposited shows strong single phase crystalline peaks in XRD (Fig 5) 2theta/omega (Fig 5 a-c , performed at (110), (100), (111) along the surface normal), clear four-fold peaks in the phi scan (Fig e-f , performed at (111), (220) along the surface normal) and rocking curve width of about a degree (Fig d). The XRD data support the growth of a near single crystal overlayer with misorientation of crystals within a degree and this XRD data is consistent at every stage of fabrication (Figure 6S, Buffer layer YSZ prior to patterning, Figure 7S buffer layer YSZ post patterning) . Further TEM studies corroborate that we have single crystal patterned YSZ on silicon (Fig 6, 8S) with mis-tilted domains (Fig 8S(c)). These mistilted domains can be attributed to the broad FWHM ( ~ 1 degrees) in the rocking curve of our crystalline thin films. This degree of crystalline misorientation is not expected to adversely affect the properties of the functional crystalline materials that would be deposited using PLATEN process.

**Discussion:** The PLATEN process does get around the challenge of patterning the complex functional oxide materials used in the fabrication of active electronic/ photonic circuitry. The process is currently limited to a lateral dimension of 50 nm, limited by our lithography. The deposited patterns exhibit smooth top and side walls with the side wall roughness limited by the silicon Bosch process. The film smoothness for crystalline case depends upon whether the film growth is dominated by layer-by-layer (smooth) or homogeneous nucleation (rough) processes.

The waist formation observed in the present PLATEN experiments is unique to the patterned substrates, while such effects are not observable on flat substrates, where the attention is more on the homogeneity of thickness, structure and composition. Furthermore, the use of alternative deposition methods, such as sputtering, e-beam and thermal evaporation, on patterned substrates results in more pronounced side wall deposition thus shadowing the waist formation. The present study implies that waist formation occurs in our samples when film thickness is more than or comparable to 80 nm. In addition, the waist width is not dependent on the width of the nano structures and is driven by the thickness (Fig 10S) with the shape of the pattern also playing a role. While we have studied the waist formation of YSZ in detail similar study for other materials is under investigation.

The fact that the amorphous low temperature deposition is so close to the high temperature deposited structures implies that to the first order the film crystallinity is not playing a dominant role in the waist formation. We suspect two competing effects:

1. The atoms forming an initial layer going beyond the edge, mediated by chemical bonding, thereby increasing the waist width.
2. The minimization of surface energy which tends to pull the film structure to reduce the surface area.

If these factors are indeed the reasons for the waist formation, we would expect the phenomenon to be independent of the material. While atomistic simulations such as large-scale molecular dynamics or Monte Carlo could provide complementary insights into grain-level phenomena, the Winterbottom construction offers an appropriate framework for film envelope morphology in systems where surface energy minimization governs shape evolution. Wulff construction was employed using computational modeling (WulffPack [22]) with Winterbottom [23] function. It is predicted that the theoretical morphology is based on the anisotropic surface energies of the various crystallographic planes in monoclinic $ZrO_2$ (Figure 9S (a)) and $ZrO_2$ onto Si (100) (Figure 9S(b)). The Wulff construction employed here provides a thermodynamic framework predicting equilibrium crystal shapes based solely on the minimization of the surface energy. Prior studies have applied Wulff construction to successfully predict thin film and nanoparticle morphologies [24], [25]. In our case, the surface energies of various m-$ZrO_2$ crystallographic planes are taken from Wolf et al [26]. Figure 9S(a) represents the equilibrium morphology of monoclinic $ZrO_2$ crystals without interfacing with a Si substrate, highlighting facets driven entirely by intrinsic surface energies. In contrast, Figure 9S(b) incorporates substrate interactions via the Winterbottom construction, illustrating how the presence of a solid-solid interface modifies the equilibrium shape, aligning with our experimental observations of directional growth transitions. The Winterbottom construction accounts for the energy balance between the film-substrate interface and the substrate-vapor interface that is replaced by this interface. In this case, the (100) surface of m-$ZrO_2$ interfaces with the Si(100), and the interfacial energy was set at 0.94 $J/m^2$. Note that the WulffPack software requires interface energies to be lower than the minimum surface energy of the crystal. Thus, the $ZrO_2$ (100) surface energy from Wolf et al. was reduced from 1.46 $J/m^2$ to 0.94 $J/m^2$ to reflect the expected similarity with the $ZrO_2$/Si interface energy, both representing oxygen-terminated surfaces with comparable bonding environments. The Wulff construction with Winterbottom function using the original surface energy from Wolf et al. is provided in the supplementary information.

It is important to emphasize that the equilibrium Wulff shape itself remains constant regardless of the crystal size, as surface energies represent intrinsic material properties that are independent of volume. However, the critical factor that evolves with increasing film thickness is the relative contribution of the substrate-film interfacial energy to the total energy of the system. For films with thicknesses below approximately 80 nm, we find that the substrate-film interfacial energy comprises a substantial fraction of the total system energy, thereby imposing significant constraints on the growth morphology and effectively preventing the crystal from maintaining its equilibrium

Wulff shape. The pattern width is invariant up to about 80 nm. As the thickness surpasses 80 nm, this interfacial contribution becomes proportionally smaller relative to the total surface energy of the crystal facets, enabling the growth to progressively transition toward the equilibrium shape predicted by our Wulffpack models. (The flat-topped morphology with width equivalent to the substrate contact area represents an energetically favorable configuration that emerges naturally as the crystal approaches its equilibrium shape). To further illustrate the thickness dependence of the predicted morphology, Winterbottom constructions were computed at five different film thicknesses 50, 100, 200, 500, and 800 nm as shown in Figure 4a and supplemental files (html). The corresponding waist heights are predicted and provided in Figure 4a. The waist height scales proportionally with thickness across this range, showing that the predicted equilibrium shape is scale-invariant and not specific to a single thickness assumption, which is consistent with the shape-preserving growth behavior observed experimentally. Furthermore, this model successfully predicts that the waist will almost always form at the center of film and reproduces the experimentally observed trend of waist height vs thickness shown in Figure 4a where it closely reproduces the high temperature deposition data. However, predicting the trend of waist width vs thickness shown in Figure 4b remains challenging as the model lacks enough boundary conditions to solve for waist width. Hence, more detailed examination is required here.

At lower deposition temperatures, a distinct morphological shift toward rounded, isotropic shapes is observed. These rounding shapes solely from the YSZ film becoming amorphous or poorly crystallized due to insufficient thermal energy during deposition, despite the crystalline Si substrate providing an epitaxial template. The reduced thermal energy limits adatom mobility on the growth surface, preventing the YSZ atoms from organizing into the well-defined crystallographic facets that would typically express the anisotropic surface energies characteristic of monoclinic YSZ. Consequently, instead of the angular faceted morphology observed at higher deposition temperatures, the low-temperature films exhibit rounded features characteristic of amorphous materials lacking long-range crystalline order. The reduced density of the film may account for the increased waist width for the lower temperature deposition case.

This waist formation phenomenon observed in depositions on patterned substrates has profound consequences from fundamental materials science to novel applications. It clearly indicates that confinement of the spatial extent of the oxides leads to their equilibrium shape being different from what one would intuitively assume based on the spatial localization imposed on the structure! How does this depend upon the ionicity or covalency of the bonds involved? How do different oxides behave, and will nitrides be any different? One other important consequence of this waist formation is the fact that at dimensions approaching sub- 50 nm, these oxides literally take a spherical shape. This essentially give us a route to creating identical nanoparticles whose dimensions can be controlled using lithography.

The process of growing well-defined crystalline thin film structure of functional oxides at nanoscale on silicon is leading us into new material growth paradigm. The YSZ buffer layer on silicon plays a pivotal role here. Currently, the highest possible quality of any (YSZ, CeO$_2$) buffer layer is only attainable on a factory grade polished silicon surface. Hence, any further surface treatment like HF etching of the oxide, hard masking during RIE or submerging the substrate in alkaline solvents for prolonged period renders the surface unsuitable for epitaxial growth. Hence, depositing a thin layer of crystalline YSZ prior to PLATEN process turned out to be a game changer due to YSZ's robustness. With further optimization we hope to improve the crystalline quality of the overlayer functional materials. Even with the current film quality, we will expect degradation of the properties of active films to be significantly low, thereby permitting useful devices.

In summary, the PLATEN process is demonstrated to be a viable way to attain patterned functional films on top of silicon with spatial resolution down to 50 nm. The growth of films on such restricted geometries allows for novel film growth mechanisms whose dependence on the atomic arrival rate, surface temperature and background oxygen pressure require further study and understanding. We believe that PLATEN will enable the rapid fabrication of optical, magnetic and electronic functional material-based circuitry on silicon enabling future electronic subsystems.

**Acknowledgement:** A.J. and P.P. acknowledge funding from the Office of Basic Energy Sciences through QuPIDC Energy Frontier Research Center under Award No. DE-SC0025620

**Figures:**

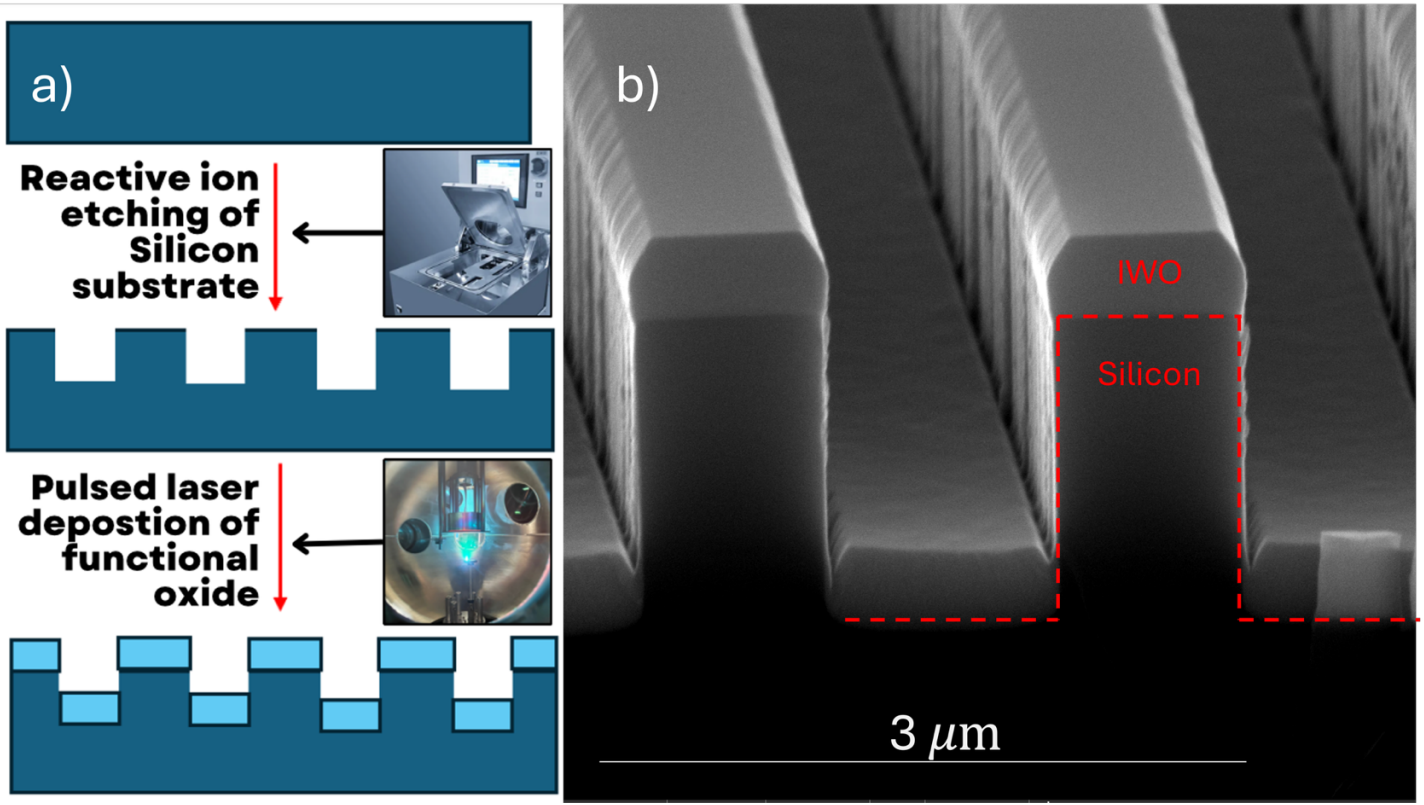

**Figure 1**: (a) Process flow of the PLATEN idea: the silicon wafer is patterned with high aspect ratio structures using the Bosch process followed by forwardly directed pulsed laser deposition of the oxide (b) 200 nm IWO film deposited at room temperature on a pre-patterned silicon substrate. The gratings have a width of 500 nm and a pitch of 1 micron.

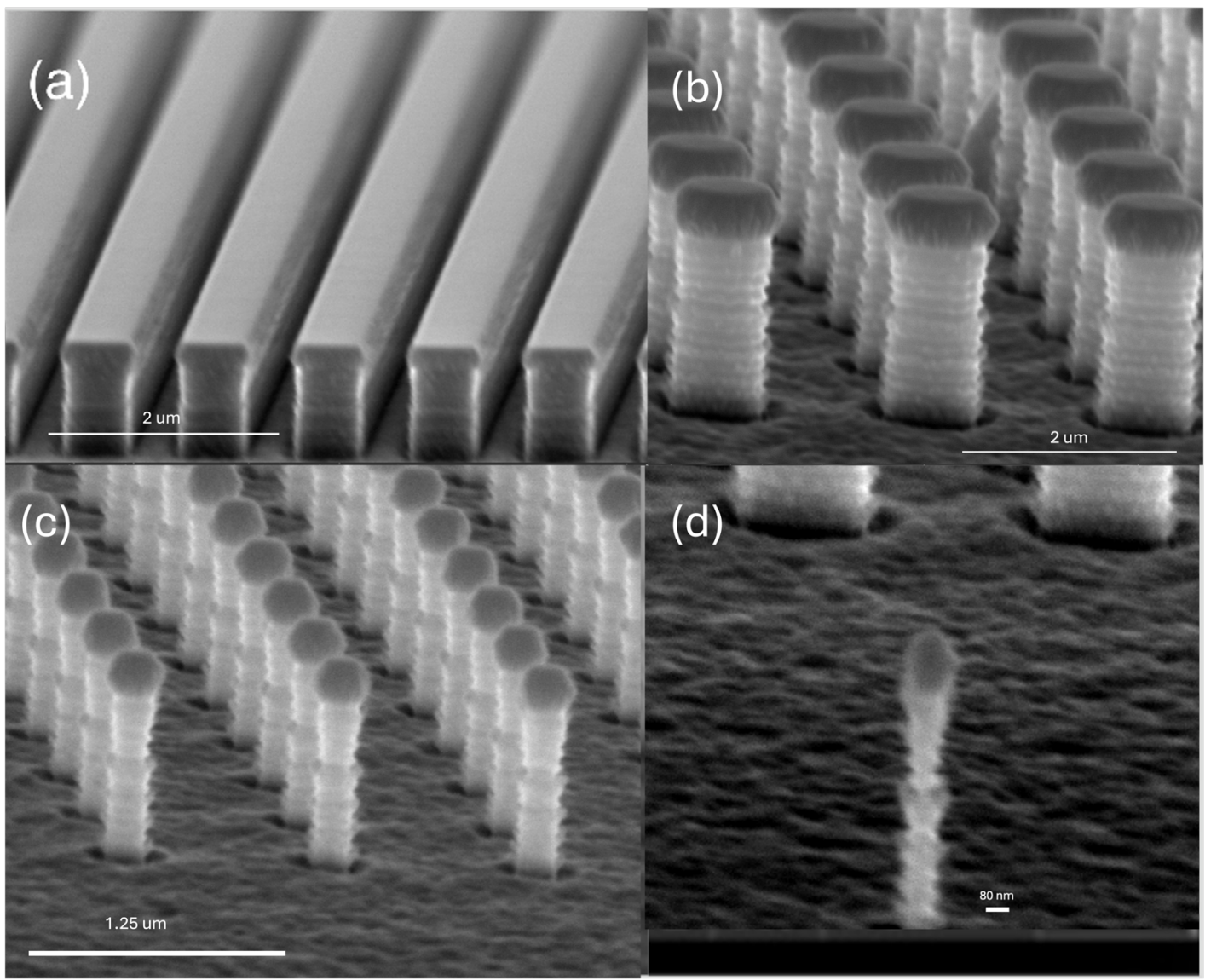

**Figure 2**: Deposition of 100 nm thick CeO$_2$ crystalline film at 800C on to pre-patterned silicon structures- (a) 500 nm wide grating, (b) 200 nm pillars, (c) 100 nm pillars and (d) < 50 nm pillar

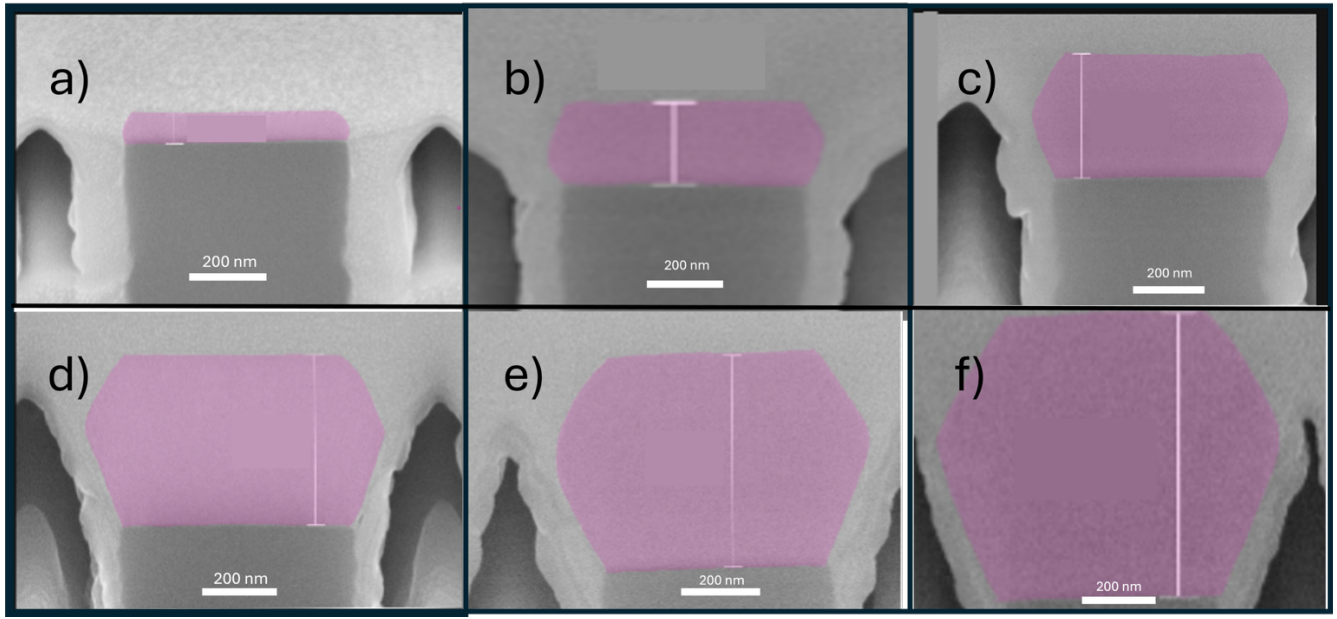

**Figure 3:** Cross-sectional SEM images of YSZ grown on silicon with various thickness at room temperature a) 74.81 nm, b) 183.1 nm, c) 320 nm, d) 420.9 nm, e) 519.2 nm f) 705.6 nm.

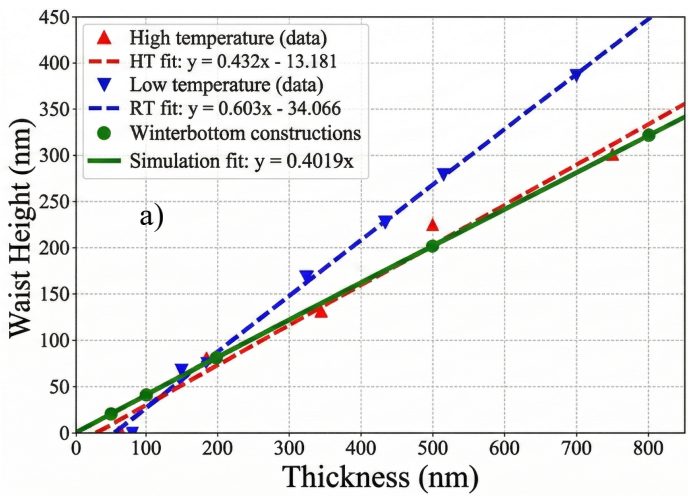
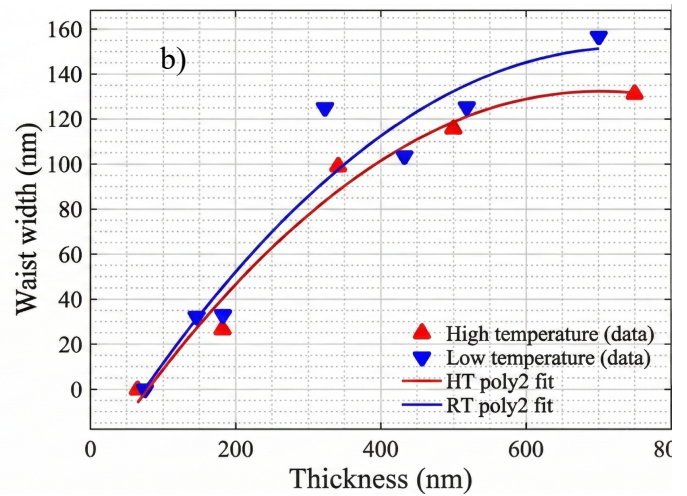

**Figure 4:** Thickness dependent growth of YSZ layer after PLATEN. The YSZ film was grown at a temperature of 950 C, with a laser energy density of 3 J/cm$^2$ and a laser repetition rate of 10 Hz. The waist forms at the middle of the film as demonstrated by the straight-line fits of 4a for high temperature deposition (red triangles), room temperature deposition (blue triangles) and the theoretical Winterbottom construction fit (green circles). 4b shows the width of the waist as a function of the film thickness.

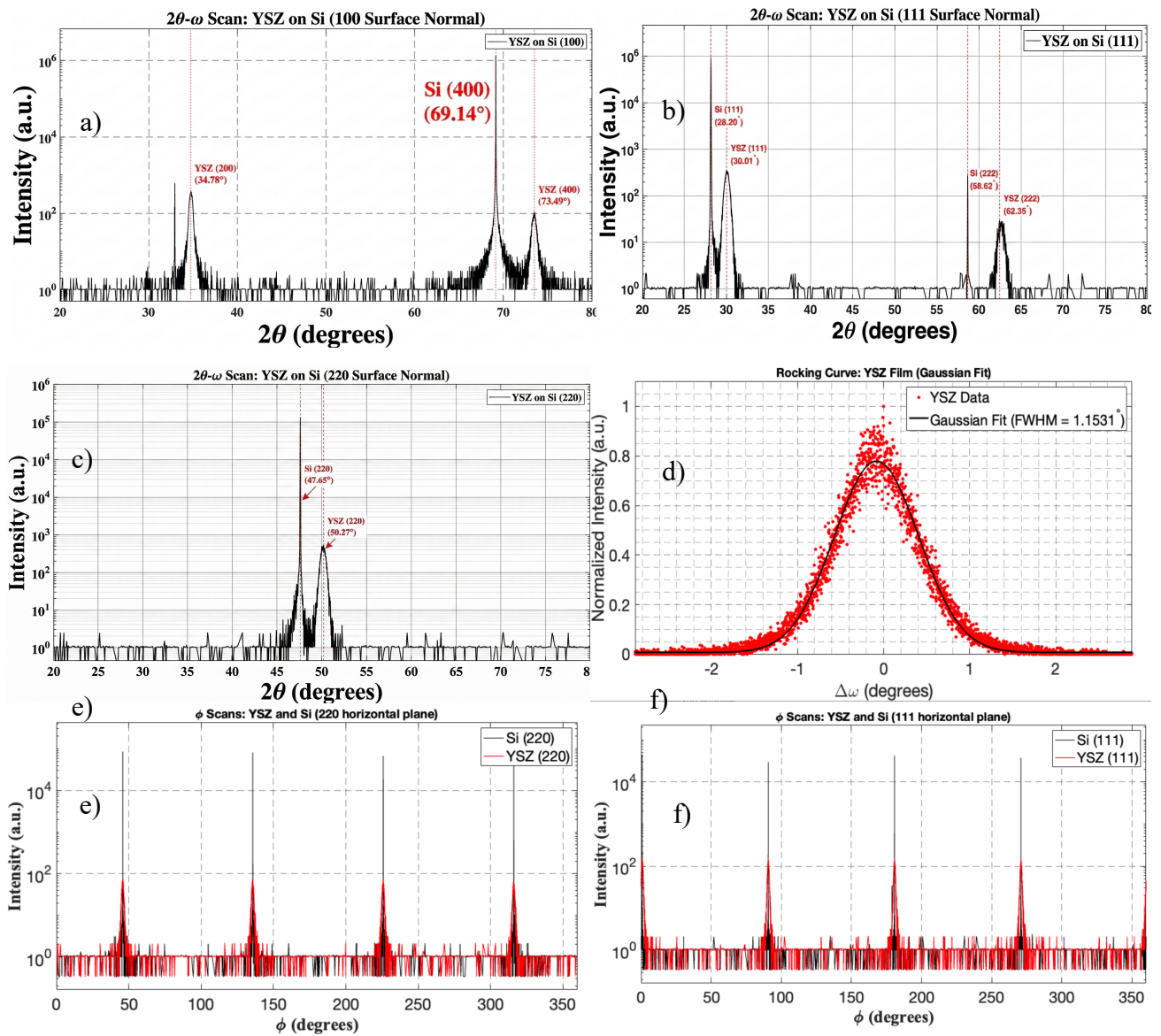

**Figure 5:** XRD analysis performed on the sample after a second layer of YSZ was deposited post patterning. 2θ scans along (a) 100, (b) 222, (c) 220 as horizontal planes proves that the film sustains its singular orientation. (d) The broad FWHM (~1.2 degrees) in the rocking curve was also unchanged and can be attributed to the presence of slight mis-tilted domains (Figure 8S(c)). Furthermore, the phi scans along (e)111, (f) 220 as horizontal planes tells us that the film's growth after patterning was homoepitaxial.

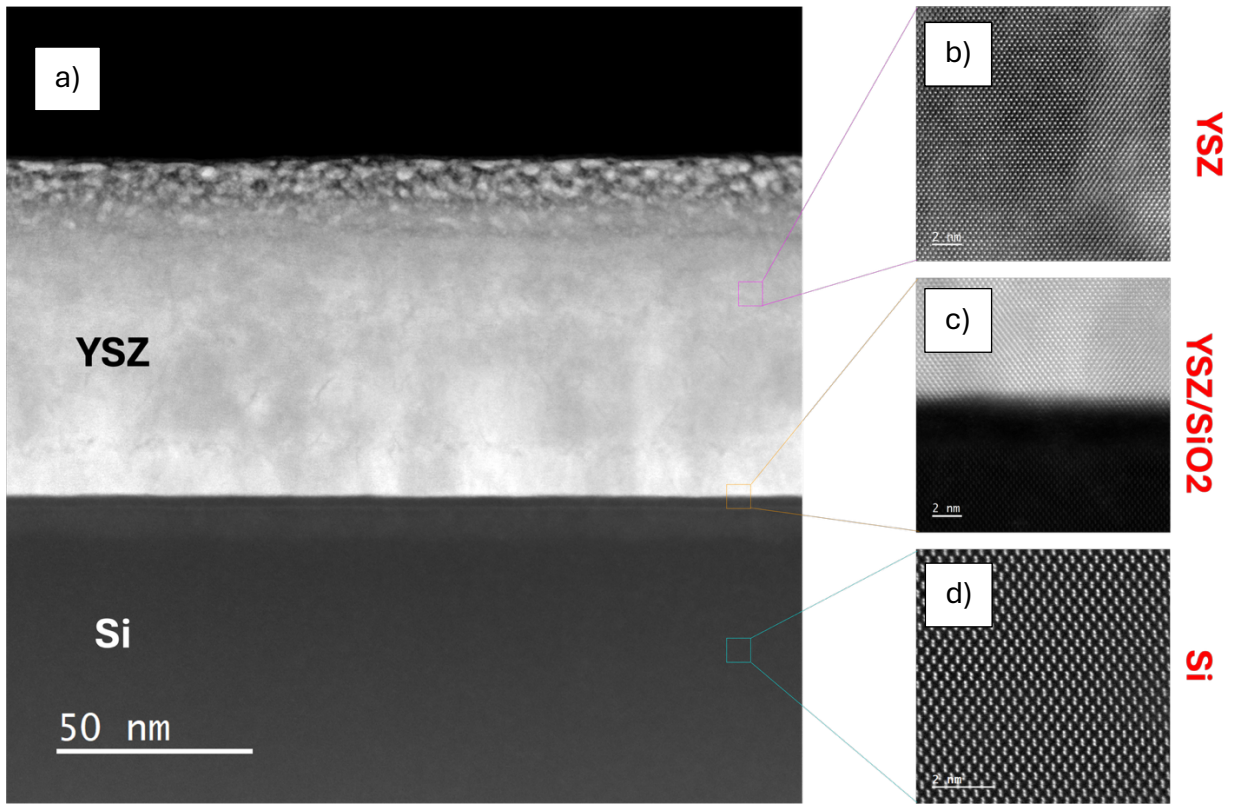

**Figure 6:** (a) Cross-sectional STEM image of YSZ grown on a patterned sample with protective YSZ layer. The YSZ film (b) deposited after patterning, (c) the protective YSZ layer and the silicon substrate interface and (d) the bottom silicon substrate are crystalline as expected.